\magnification =\magstep1
\baselineskip = 24 true pt
\hsize= 15 true cm
\vsize = 22 true cm
\centerline{Expulsion   of   Magnetic  Flux  Lines  from  Growing
Superconducting}
\centerline{Core of a Quark Star and The Possibility of }
\centerline{Mullins-Sekerka Interface Instability}

\bigskip
\centerline{Somenath Chakrabarty}
\medskip

\centerline{\footnote\dag{Permanent Address} Department of Physics,
University of Kalyani}
\centerline{West Bengal, India 741 235}
\centerline{E-Mail:somenath@klyuniv.ernet.in}
\centerline{and}
\centerline{Inter-University centre for Astronomy \& Astrophysics}
\centerline{Post Bag 4, Ganeshkhind}
\centerline{Pune 411 007, India}

\bigskip\bigskip\bigskip

\centerline{Abstract}

\bigskip

\noindent  The  expulsion  of  magnetic flux lines from a growing
superconducting core of a quark star has been investigated  using
the  idea  of impurity diffusion in molten alloys. The possibility
of Mullins-Sekerka normal-superconducting  interface  instability
has also been studied.

\vfil

\noindent PACS NO.:24.85+p, 97.60Jd, 74.20Hi, 64.70-p

\eject

If the matter density at the core of a neutron star exceeds a few
times    normal    nuclear    density    (e.g.   $>3n_0$,   where
$n_0=0.17$fm$^{-3}$, the normal nuclear density),  a  deconfining
phase transition to quark matter may take place. As a consequence
a  normal neutron star will be converted to a hybrid star with an
infinite cluster of quark matter core  and  a  crust  of  neutron
matter. If the speculation of Witten [1] that a flavour symmetric
quark matter may be the absolute ground state at zero temperature
and  pressure is true, there is a possibility that the whole star
will be converted to a star of strange quark matter  (SQM)  known
as  strange  star.  In normal quark matter the strange quarks are
produced through  flavour  non-conserving  weak  processes  which
ultimately lead to a dynamical chemical equilibrium among various
constituents.

From  the observed features in the spectra of pulsating accreting
neutron stars in binary system, the strength of surface  magnetic
field  of  a  neutron star is found to be $\sim 10^{12}$G. At the
core region of a newly born  neutron  star  it  probably  reaches
$\sim  10^{18}$G  [2].  In a recent publication [3] we have shown
that if the magnetic field intensity exceeds some critical  value
which  is  the  typical  strength  of magnetic field at which the
cyclotron lines begin to  occur  or  equivalently  at  which  the
cyclotron  quantum  is  of  the order of or greater than the rest
mass of the particle considered or the de Broglie wave length  is
of  the  order  of  or  greater  than  the  Larmor  radius of the
particle, there can not be nucleation of a  single  quark  matter
bubble  in  the metastable neutron matter. The surface as well as
the curvature energies diverge in this case. As a consequence the
new stable (quark matter)  phase  can  not  be  thermodynamically
favourable  over  metastable (neutron matter) phase. Therefore to
achieve a first order deconfining  transition  initiated  by  the
nucleation  of  quark  droplets  at  the core of neutron star, we
assume that the strength of magnetic field throughout the star is
much less than the corresponding critical value. In the  case  of
electron of mass $0.5$MeV, this critical field is $\sim 4.4\times
10^{13}$G,  for  light quarks of current mass $5$MeV, it is $\sim
10^{15}$G, whereas for $s$-quark of current mass $150$MeV, it  is
$\sim 10^{20}$G [4-6].

Now  for  a  many  body fermion system, the microscopic theory of
superconductivity suggests [7] that if  the  interaction  favours
formation  of  pairs at low temperature, the system may undergo a
phase transition to a superfluid state. This is expected to occur
in the dense neutron matter present in neutron star [8,9]. On the
otherhand, if the particles carry charges, the paired state  will
be  superconducting.  In  the  case  of  an electronic system BCS
theory applied to study the superconducting properties  [7].  One
electron  of  momentum  $\vec  k$ and spin $\vec s$ combines with
another one of momentum $-\vec k$ and spin $-\vec s$ and  form  a
Cooper  pair.  The  coupling  is  mediated by the electron-phonon
interaction in the  lattice.  In  the  case  of  SQM,  the  basic
quark-quark  interaction  is  attractive  at  large  distance and
consequently the BCS pairing mechanism is also  applicable  here.
For  a  highly  degenerate system, which is true in strange star,
the pairing  takes  place  near  the  fermi  surface.  The  other
condition  that must be satisfied to form Cooper pair is that the
temperature ($T$) of the system should  be  much  less  than  the
superconducting energy gap ($\Delta$), which is a function of the
interaction  strength  and the density of the system. This is the
most important criterion for  the  occurance  of  superconducting
transition.  In  the  case  of strange star, only quarks can form
Cooper pairs. The electrons, whose density is extremely low, form
highly degenerate  relativistic  plasma,  are  unlikely  to  form
Cooper pairs. The kinetic energy of the electronic part dominates
over  its  attractive  potential  energy,  and  as  a  result the
corresponding superconducting transition temperature is extremely
low and may not be achieved in a strange star.  The  relativistic
theory  of  superfluidity  and  superconductivity  for  a fermion
system was given by Bailin and Love [10]. Recently, Horvath et al
[11] and also we have studied [12] the superconductivity of quark
matter using the results of  ref.  [10].  We  have  also  studied
qualitatively  the  magnetic  properties  of quark matter in ref.
[12] for massive quarks.

Now  the  quarks  of  same  fermi energy can only combine to form
Cooper pairs. Since the $u$ and $d$ current masses are equal  and
also  their chemical potentials are almost identical, whereas $s$
quark is much heavier than  $u$  and  $d$  quarks  and  also  its
chemical  potential  is  different,  we can have only $uu$, $dd$,
$ud$ and $ss$ Cooper pairs in  the  system.  For  iso-spin  $1/2$
flavours,  the  contribution  may  come either from iso-scalar or
iso-vector channels. It was shown in ref. [10] that  the  pairing
of  a  $u-d$  system  will  be favoured by iso-scalar combination
rather than iso-vector channel.  On  the  other  hand  the  $s-s$
combination  is  a  triplet  state  with  $J^P=1^+$. Now from the
conclusions of refs. [10-12] we know that if a normal SQM  system
undergoes  a superconducting phase transition, the newly produced
SQM phase will be a type I superconductor. We have also seen that
the critical magnetic field for such pairing is  $\sim  10^{16}$G
for  $n\sim 2-3n_0$, which is indeed much larger than the typical
pulsar magnetic field. The corresponding critical temperature  is
$10^9-10^{10}$K,  this  can  possibily be achieved in quark star,
which is expected to be an extremely cold object. Since the quark
star magnetic field is less than the corresponding critical field
for the  destruction  of  superconductivity,  during  such  phase
transition,  the  magnetic  flux  lines  from the superconducting
quark sector of the strange star will be pushed out  towards  the
normal  crust  region.  Unlike  a  small  type  I superconducting
laboratory sample placed in an external magnetic field (<critical
value) for which the expulsion  of  magnetic  field  takes  place
almost  instantaneously, in this particular case the scenario may
be completely different.

The  aim  of the present note is to investigate using the idea of
impurity diffusion in molten alloys, the  expulsion  of  magnetic
flux  lines  from growing superconducting core of a strange star.
We  have  also  studied  the   possibility   of   Mullins-Sekerka
normal-superconducting   interface   instability  [13]  in  quark
matter.  This  is  generally  observed  (i)  in   the   case   of
solidification   of   pure  molten  metals  at  the  solid-liquid
interface, if there is a temperature gradient. The interface will
always be stable if the  temperature  gradient  is  positive  and
unstable  otherwise, (ii) during solidification of molten alloys.
In alloys, the criteria for stable / unstable behaviour  is  more
complicated.  It  is  seen  that, during the solidification of an
alloy, there is a substantial change in the  concentration  ahead
of  the interface. Here solute diffusion as well as the heat flow
effects must be considered simultaneously. As we  will  see,  the
particular  problem  we are going to investigate is analogous to
solute diffusion during solidification of an alloy.

It  has  been  assumed  that  the growth of superconducting quark
bubble has started from the centre of the star  and  we  use  the
nomenclature {\it{controlled growth}} for such phenomenon. If the
magnetic  field strength and the temperature of the star are much
less  than  their  critical  values,  the  normal  SQM  phase  is
thermodynamically   unstable   relative   to   the  corresponding
superconducting one.  Then  due  to  fluctuation,  a  droplet  of
superconducting quark matter bubble may be produced in metastable
medium.  If  the  size  of this superconducting bubble is greater
than the  corresponding  critical  value,  it  will  act  as  the
nucleating  centre  for the growth of superconducting quark core.
The critical radius can be obtained by minimizing the free energy
and                 is                  given                  by
$r_c=16\pi\alpha/B_m^{(c)2}[1-(B_m/B_m^{(c)})^2]$, where $\alpha$
is  the  surface tension, which is greater than zero for a type I
superconductor-normal  interface,  $B_m^{(c)}$  is  the  critical
magnetic field. In presence of a magnetic field $B_m< B_m^{(c)}$,
the  normal  to  superconducting  transition  is  first  order in
nature. As the  superconducting  phase  grows  continuously,  the
magnetic  field  lines  will  be pushed out into the normal quark
matter crust. This is the usual {\it {Meissner effect}}, observed
in type I superconductor. We compare this phenomenon of  magnetic
flux  expulsion  from a growing superconducting SQM core with the
diffusion of impurities from a molten  metal.  The  formation  of
superconducting  zone  is  compared  with  the  solidification of
molten metal. It is known from simple thermodynamic  calculations
that  if the free energy of molten phase decreases in presence of
impurity atoms, then during solidification they prefer to  recide
in  the  molten phase. In this particular case the magnetic field
lines play the role of impurity atoms, the  normal  quark  matter
phase plays the role of molten metal, whereas the superconducting
phase can be compared with the frozen solid phase. (This idea was
recently  applied  to  baryon number transport during first order
quark-hadron phase transition in the early Universe, where baryon
number replaces impurity, quark phase replaces molten  metal  and
hadronic  matter replaces that of solid metal [14]). The magnetic
flux lines prefer  to  recide  in  the  normal  phase.  Then  the
Meissner  effect  can  be  restated  as  {\it{the  solubility  of
magnetic flux lines in the superconducting phase  is  zero}}  (of
course, there is a finite penetration depth).

The  dynamical  equation  for  the flux expulsion can be obtained
from  the  simplified  model  of   sharp   normal-superconducting
interface.  The  expulsion  equation  is  given by the well known
diffusion equation [15]
$${{\partial B_m}\over {\partial t}}=D\nabla^2 B_m\eqno(1)$$
where $B_m$ is the  magnetic  field  intensity  and  $D$  is  the
diffusion coefficient, given by
$$D={{c^2}\over{4\pi \sigma_n}}\eqno(2)$$
where $\sigma_n$ is the electrical conductivity of the normal SQM
phase,  for  superconducting  phase $B_m=0$. Following ref. [16],
the electrical conductivity of SQM for $B_m=0$ is given by
$$\sigma_n=5.8\times  10^{25} \left ({{\alpha_c}\over{0.1}}\right
)^{-3/2} T_{10}^{-2} \left ( {{n}\over{n_0}}\right ) \eqno(3)$$
in  sec$^{-1}$,  where $\alpha_c$ is the strong coupling constant
and   $T_{10}=T/10^{10}$K,   the   value   of   this   electrical
conductivity in  the  case  of  strange  quark  matter  is  $\sim
10^{26}$ sec$^{-1}$. We have used this expression to get an order
of magnitude estimate of electrical conductivity of quark matter.
In  actual calculation one has to evaluate $\sigma_n$ in presence
of $B_m$. In that case, $\sigma_n$ can not be a scalar  quantity.
In  particular,  for extremely  large  $B_m$,  the components orthogonal to
$B_m$  tend to zero. The quarks can only move along the direction
of magnetic field, across the field resistivity becomes infinity.

A solution of eqn.(1) with spherical symmetry (which may not be a
valid assumption) can be obtained by Green's function  technique,
and  is  given  by  (for a very general topological structure, no
analytical solution is possible)
$$B_m(r,t)={{1}\over{2r (\pi D t)^{1/2}}} \int_0^\infty B_m^{(0)}
(r') \big [ e^{-u_-^2}-e^{-u_+^2}\big ] r'dr'\eqno(4)$$
where  $u_\pm=(r\pm  r')/2(Dt)^{1/2}$  and  $B_m^{(0)}(r)$ is the
magnetic field distribution within the star at $t=0$, which is of
course an entirely unknown function of radial coordinate $r$.  To
obtain an idea of magnetic field diffusion time scale ($\tau_D$),
we assume $B_m^{(0)}(r)=B_m^{(0)}=$ constant (in reality, this is
not  possible  inside the star). Then using the other approximate
result for electrical  conductivity  (which  is  valid  for  zero
magnetic    field    case),    given    by   eqn.(3),   we   have
$\tau_D=10-20$  yrs.  With  this   constant   $B_m^{(0)}(r)$,
eqn.(4)
$$B_m(r,t)=B_m^{(0)}  \left  [  {{1}\over{2}} \left \{ {\rm{erf}}(u_+)
+{\rm{erf}}(u_-)\right \} +{{1}\over{r}} \left  ({{Dt}\over{\pi}}\right
)^{1/2} \left \{ e^{-u_+^2}-e^{-u_-^2}\right \} \right ]\eqno(5)$$
where
$${\rm{erf}}(x)={{2}\over{(\pi)^{1/2}}} \int_0^x e^{-z^2}dz$$
is the well known error function.

From the eqn.(5), using the approximate  results  for  electrical
conductivity,  given  by eqn.(3) (which is valid for $B_m=0$) one
can get an estimate of time scale for the expulsion  of  magnetic
lines  of  force and is $\sim 10-20$ yrs. Latter we shall see that
such a time scale can also be obtained from stability analysis of
planer   normal-superconducting   interface.   Again   the   time
dependence of the  electrical  conductivity  profile  for  normal
quark  matter  is  not  known.  This  is  another  uncertainty in
obtaining exact solution for the diffusion equation.

Therefore, almost nothing can be said about the growth of superconducting
zone and the expulsion of magnetic flux lines from this region by
solving  eqn.(1).  We shall now study the morphological
instability of normal-superconducting interface of  quark  matter
in the star using the idea of
solute diffusion during solidification of alloys. The  motion  of
normal-superconducting  interface  is extremely important in this
case and has
to be taken into consideration. Then instead of eqn.(1) which  is
valid  in  the  rest frame, an equation expressed in a coordinate
system which is moving with an element of the boundary  layer  is
the  correct description of such superconducting growth, known as
{\it{Directional   Grwoth}},   and   the   equation   is   called
{\it{Directional Growth Equation}}, and is given by
$${{\partial  B_m}\over  {\partial  t}}  -v  {{\partial B_m}\over
{\partial z}} =D\nabla^2 B_m\eqno(6)$$
where  the motion of the interface is along the z-axis and $v$ is
the velocity of  the  front.  This  diffusion  equation  must  be
supplemented  by  the  boundary  conditions at the interface. The
first boundary condition is obtained by  combining  Ampere's
and Faraday's laws at the interface, and is given by
$$B_mv\mid_s=-D(\nabla B_m) .\hat n \mid_s \eqno(7)$$
where   $\hat   n$  is  the  unit  vector normal to the interface
directed from the normal phase to the superconducting phase. This
is actually the continuity equation for magnetic flux diffusion. The  rate
at  which  excess  magnetic  field  lines  are  rejected from the
interior of the phase is balanced by the rate at  which  magnetic
flux lines diffuses ahead of the two-phase interface. This effect
makes  the  boundary  layer  between superconducting-normal quark
matter phases unstable due to excess magnetic field lines present
on the surface  of  the  growing  superconducting  bubble.  Local
thermodynamic  equilibrium at the interface gives (Gibbs-Thompson
condition)
$$B_m \mid_s  \approx  B_m^{(c)}  \left  (  1-{{4\pi   \alpha
}\over{RB_m^{(c)^2}}}\right )
=  B_m^{(c)}  \left  (  1- \delta c \right ) \eqno(8)$$
where  $\delta$  is  called  capillary  length  with $\alpha$ the
surface tension, $c$ is the curvature  $=1/R$  (for  a  spherical
surface), and $B_m^{(c)}$ is the thermodynamic critical field.

To investigate the stability of superconducting-normal interface,
we  shall  follow  the original work by Mullins and Sekerka [13],
and consider a steady state growth of superconducting core,  then
the  time  derivative  in  eqn.(6)  will  not appear. Introducing
$r_\perp=(x^2+y^2)^{1/2}$ as the transeverse  coordinate  at  the
interface, we have after rearranging eqn.(6)
$$\left      [{{\partial^2      }\over{\partial      r_\perp^2}}+
{{1}\over{r_\perp}} {{\partial}\over{\partial r_\perp}}+
{{\partial^2 }\over{\partial z^2}}+
{{v}\over   {D}}{{\partial   }\over{\partial  z}}\right  ]  B_m=0
\eqno(9)$$
Following the most common approximation which is made in the case
of  freezing  of  molten  solid  is  that  the  solidification is
occuring under steady state condition, which in  this  particular
case is the normal to superconducting phase transition, and that,
therefore,  the  concentration  of  magnetic   flux   lines   and
normal-superconducting  interface  morphology  are independent of
time. The principal disadvantage of this assumption  is  that  no
evolution  of  the  interface shape can occur. The result of this
constraint is that the solution  to the basic  diffusion  problem
is   indeterminate   and   a   whole  range  of  morphologies  is
permissible from the mathematical point  of  view.  In  order  to
distinguish  the  solution which is the most likely to correspond
to reality, it is necessary to  find  some  additional  criteria.
Examination  of  the  stabilities  of a slightly perturbed growth
form is probably the most reasonable manner  in  which  to  treat
this  situation.  In  the  following  we  shall  investigate  the
morphological  instability  of  normal-superconducting  interface
following eqn.(9).
Assuming  a solution of this equation expressed as the product of
separate  functions  of  $r_\perp$  and  $z$  and   setting   the
separation  constatnt  equal  to  zero and  using the boundary
condition given by eqn.(8), we have for an  unperturbed  boundary
layer moving along $z$-axis
$$B_m=B_m^{(s)} e^{-zv/D} =B_m^{(s)} e^{-2z/l}\eqno(10)$$
where $l=2D/v$ is the layer thickness, which is  very  small  for
small $D$. Mthematicaly, the thickness of this layer is infinity.
For practical purpose an effective value $l$ can  be  taken.  The
order  of  magnitude  estimates  or limiting values for the three
quantities $D$, $v$ and $l$ can be obtained  from  the  stability
condition of planer interface, which will be discussed latter.

Due  to  excess magnetic flux lines at the interface, the form of
the planer  normal-superconducting  interface  described  by  the
equation  $z=0$  is assumed to be changed by a small perturbation
represented by the simple sine function
$$z=\epsilon \sin(\vec k.\vec r_\perp)\eqno(11)$$
where  $\epsilon$  is  very  small  amplitude and $\vec k$ is the wave
vector of the perturbation. Then the perturbed  solution  of  the
magnetic field distribution near the interface can be written as
$$B_m=B_m^{(s)}e^{-vz/D}      +A\epsilon     \sin(\vec     k.\vec
r_\perp)e^{-bz}\eqno(12)$$
where  $A$  and $b$ are two unknown constants. Since the solution
should satisfy the diffusion equation (9), we have
$$b={{v}\over{2D}}+\left     [     \left    ({{v}\over{2D}}\right
)^2+k^2\right ]^{1/2} \eqno(13)$$
To  evaluate  $A$,  we utilise the assumption that $\epsilon$ and
$\epsilon \sin(\vec k.\vec r_\perp)$ are small enough so that  we
can  keep  only the linear terms in the expansion of exponentials
present  in  eqn.(12). Then after some straight forward algebraic
manipulation, we have
$$A=-{{v}\over{D}} B_m^{(s)}\eqno(14)$$
The  expression  describing the magnetic field distribution ahead
of the slightly perturbed interface then reduces to
$$B_m=B_m^{(s)} \left [ e^{-vz/D}  -{{v}\over{D}}  \epsilon  \sin
(\vec k.\vec r_\perp)e^{-bz} \right ]\eqno(15)$$

Now from the other boundary condition (eqn.(8)) we have
$$B_m^{(s)}=B_m^{(c)}-{{4\pi \alpha B_m^{(c)}}\over{B_m^{(s)2}}}c
\eqno(16)$$
where $c=z^{''}/(1+z^{'2})^{3/2}$ is the curvature at $z=\epsilon
\sin(\vec  k.\vec  r_\perp)$  and prime indicates derivative with
respect to $r_\perp$.

Neglecting $z^{'2}$, which is small for small perturbation, we have
$$B_m^{(s)}=B_m^{(c)} +\Gamma k^2 S\eqno(17)$$
where $\Gamma =4\pi  \alpha  B_m^{(c)}/B_m^{(s)2}$  and  we  have
replaced  $\epsilon  \sin(\vec k.\vec r_\perp)$ by $S$. Since the
amplitude of perturbation  $\epsilon$  is  extremely  small,  the
quantity $S$ is also negligibly small.

Now the eqn.(17) is also given by
$$B_m^{(s)}= B_m^{(c)}+ GS \eqno(18)$$
where
$$G={{dB_m}\over{dz}}\mid_{z=S}     =-{{v}\over{D}}    \left    (
1-{{vS}\over{D}} \right ) B_m^{(s)} -bAS (1-bS) \eqno(19)$$
Combining these two eqns., we have
$$k^2\Gamma +{{v}\over{D}}  \left  (  1-{{vS}\over{D}}  \right  )
B_m^{(s)} -{{bv}\over{D}} B_m^{(s)} S(1-bS)=0\eqno(20)$$
This  expression determines the form  (values of $k$) which the
perturbed interface must assume in order to satisfy  all  of  the
conditions  of the problem. To analyse the behaviour of the roots
we replace right hand side of eqn.(20) by  some  parameter  $-P$.
(We  have  taken  $-P$  in order to draw a close analogy with the
method given in ref.[13]). Rearranging eqn.(20), we have
$$-k^2\Gamma  -{{v}\over{D}}  \left  (  1-{{vS}\over{D}} \right )
B_m^{(s)} +{{bvB_m^{(s)} S}\over{D}} (1-bS) =P\eqno(21)$$
(in  ref.  [13]  the  parameter  $P$  is  related  to  the time
derivative $\epsilon$ of the amplitude of small perturbation). If
the  parameter  $P$  is  positive  for  any  value  of  $k$,  the
distortion  of  the  interface  will  increase,  whereas, if it is
negative for all values of $k$, the perturbation  will  disappear
and  the interface will be stable. In order to derive a stability
criterion, one only needs to know whether eqn.(21) has roots  for
positive values of $k$. If it has no roots, then the interface is
stable  because  the  $P-k$  curve never rises above the positive
$k$-axis and $P$ is therefore negative for  all  wavelengths.  We
have  used Decarte's theorem to check how many positive roots are
there. It is more convenient to express $k$ in terms of  $b$  and
then replacing $b$ by $\omega +v/D$, which gives
$$-\omega^2 \left ( \Gamma +{{vB_m^{(s)} S^2}\over{D}}  \right  )
-\omega  \left ( \Gamma +{{2vB_m^{(s)} S^2}\over{ D}} -B_m^{(s)} S
\right  )  {{v}\over{D}}  -{{v}\over{D}}  B_m^{(s)}  \left  (  1-
{{v}\over {D}} S\right )^2 =P\eqno(22)$$
This  is  a  quadratic equation for $\omega$. The first and the
third terms are always negative. The second  term  will  also  be
negative if
$$\Gamma + {{2vB_m^{(s)} S^2}\over{D}} -B_m^{(s)} S>0\eqno(23)$$
Then  it follows from Decart's rule that if the condition (23) is
satisfied, there can not be any positive root. Which implies that
the small perturbation of the interface will disappear. Since the
amplitude of perturbation is assumed to be extremely  small,  the
quantity   $S=\epsilon   \sin(\vec  k  .\vec  r_\perp)$  is  also
negligibly small. Under such circumstances  the  middle  term  of
eqn.(23)  is  much  smaller  than rest of the terms. The Decart's
rule given by the condition (21) can be re-written as
$$\Gamma > B_m^{(s)} S\eqno(24)$$
Which after some simplification gives the stability criterion for
the plane unperturbed interface, given by
$$\alpha >{{B_m^{(s)3}S}\over{4\pi B_m^{(c)}}}\eqno(25)$$

From   the   stability   criterion,   it   follows    that    the
normal-superconducting  interface energy/area of quark matter has
a lower bound, which depends  on  the  interface  magnetic  field
strength,  critical  field  strength and also on the perturbation
term $S$. An order  of  magnitude  of  this  lower  limit  can  be
obtained by asuming $B_m^{(s)}=10^{-3}B_m^{(c)}$.
(Since  the  critical  field  $B_m^{(c)}\sim  10^{16}$G,  and the
neutron star magnetic field strength $B_m\sim 10^{13}$G,  we  may
use this equality). Then the lower limit is given by
$$\alpha_L\approx  10^{-9}~~{\rm{MeV/fm}}^2~\left( {{S}\over{{\rm
{fm}}}}\right )\eqno(26a)$$
On the other hand for $B_m^{(s)}=0.1 B_m^{(c)}$, we have
$$\alpha_L\approx  10^{-3}~~{\rm{MeV/fm}}^2~\left( {{S}\over{{\rm
{fm}}}}\right )\eqno(26b)$$
The approximate general expression for the lower limit  is  given
by
$$\alpha_L\approx  h^3~{\rm{MeV/fm}}^2~\left( {{S}\over{{\rm
{fm}}}}\right )\eqno(26c)$$
where  $h=B_m^{(s)}/B_m^{(c)}$.  There  for  the maximum value of
this lower limit is
$$\alpha_L^{\rm{max.}}\approx  1~{\rm{MeV/fm}}^2~\left( {{S}\over{{\rm
{fm}}}}\right )\eqno(26d)$$
when the two phase are in thermodynamic  equilibrium.  Of  coures
for  such  a strong magnetic field, as we have seen (see ref.[3])
that there can not be first order quark-hadron phase  transition,
and it should be some higher order one.

On  the  other  hand  if  we do not have control on the interface
energy, which can in principle be obtained  from  Landau-Ginzberg
model,   we  can  re-write  the  stability  criteria  interms  of
interface concentration of magnetic field  strength  $B_m^{(s)}$,
and is given by
$$B_m^{(s)}  < \left [ {{4\pi \alpha B_m^{(c)} }\over{S\left ( 1-
{{2v}\over{D}}S\right )}} \right ]^{1/3}\eqno(27)$$
This  is more realistic than the condition imposed on the surface
tension $\alpha$. Now for a type I  superconductor,  the  surface
tension  $\alpha  >0$,  which implies $1-2vS/D >0$. Therefore, we
have  $2vS/D  <  1$,  and  for  the   typical   value   $\sigma_n
\sim10^{26}$ sec$^{-1}$ for the electrical conductivity of normal
quark  matter  with zero magnetic field, the profile velocity $v<
D/2S \sim 10^{-6}/S$ cm/sec $\sim 1$ cm/sec for  $S\sim  10^{-6}$
cm. Therefore the interface velocity $<1$ cm/sec for such typical
values  of $\sigma_n$ and $S$ to make the planer interface stable
under small perturbation. Now the thickness of the layer  at  the
interface   is  $l=2D/v>10^{-6}$  for  such  values  of  $D$  (or
$\sigma_n$)  and  $v$.  Here  $S$  is  always  greater  than  $0$,
otherwise,     the     magnetic    field    strength    at    the
normal-superconductor interface becomes unphysical. As before, if
the second term of eqn.(23) is negligibly small compared to other
two terms, we have
$$B_m^{(s)}  <  \left  [  {{4\pi \alpha B_m^{(c)}}\over{S}}\right
]^{1/3} \eqno(28)$$

Therefore  we  may  conclude  by saying that if a superconducting
tnasition can take place in a quark star, the magnetic properties
of such bulk object are entirely different from that of  a  small
laboratory  superconducting  sample.  Expulsion  of magnetic flux
lines from the superconducting zone  is  not  instantaneous.  The
typical  time scale is $10-15$ yrs. Due to the presence of excess
magnetic flux lines at the interface, which is possibly  true  if
the  diffusion  rate  of  magnetic  lines of forces in the normal
phase is less than the rate  of  growth  of  the  superconducting
zone, the characteristic of normal-superconducting boundary layer
may  change significantly. Of course, it depends on the magnitude
of surface tension $\alpha$. It may take dendritic shape  instead
of  planer  structure.  The  stability  of  planer interface also
depends on the strength of interface magnetic field, if we do not
have control on the interface energy and are given  by  eqns.(27)
and (28). How to get an experimental evidence for such an unusual
shape is a matter of further study.
\vfil\eject

\noindent References

\item{1.} E. Witten, Phys. Rev. D30, 272 (1984).
\item{2.}  J.  Trumper  et  al,  Ap.  J.  219, L105 (1978); W. A.
Wheaton et al, Nature 272, 240 (1979); D. E. Gruber et al, Ap. J.
240, L127 (1980); T. Mihara et al, Nature 346, 250 (1990).
\item{3.} S. Chakrabarty, Phys. Rev. D51, 4591 (1995).
\item{4.}  S.  Chakrabarty,  Astrphys.  and  Space  Sci., 213, 121
(1994).
\item{5.} S. Chakrabarty and A. Goyal, Mod. Phys. Lett. A9,  3611
(1994).
\item{6.}  S.  Chakrabarty and P. K. Sahu, Phys. Rev. D (in press
1996); S. Chakrabarty, Phys. Rev. D (submitted).
\item{7.}  A.L.  Fetter and J. D. Walecha, Quantum Theory of Many
Particle System, McGraw Hill Book Company, New York, 1971.
\item{8.} G. Baym, C. Pethick  and  D.  Pines,  Nature  224,  224
(1969).
\item{9.} M. Baldo et al, Nucl. Phys. A536, 349 (1992).
\item{10.} D. Bailin and A. Love, Phys. Rep. 107, 325 (1984).
\item{11.} J. E. Horvath et al, Mod. Phys. Lett. A7, 995 (1992).
\item{12.} S. Chakrabarty, Can. J. Phys. 71, 488 (1993).
\item{13.}  W.W. Mullins and R. F. Sekerka, Jour. Appl. Phys. 34,
323 (1963); 35, 444 (1964).
\item{14.} F. C. Adams, K. Freese and  J.  S.  Langer, Phys. Rev.
D47, 4303 (1993);  Marc Kamionkowski and K. Freese, Phys.
Rev. Lett. 69, 2743 (1992).
\item{15.} J. S. Langer, Rev. Mod. Phys.  52,  1  (1980).
\item{16.}  P. Haensel and A. J. Jerzak, Acta Phys. Pol. B20, 141
(1989); P. Haensel, Nucl. Phys. B24 (proc. of the  Int.  Workshop
on  Strange  quark  Matter  in Physics and Astrophysics, Univ. of
Aarhus, Denmark, May 20-24, 1991), 23 (1991). \vfil\eject\end

\vfil\eject\end